\documentclass[prb, aps, bibliography, twocolumn, reprint, showpacs, footnoteinbib,superscriptaddress]{revtex4-1}

\usepackage{graphicx}
\usepackage{amssymb}   
\usepackage{amsfonts}
\usepackage{amsmath}
\usepackage{hyperref}
\usepackage{color}
\usepackage{braket}
\usepackage{overpic}
\usepackage{rotating}
\newcommand{\be}{\begin{equation}}
\newcommand{\ee}{\end{equation}}

\begin{document}
	
\title{Measuring complex partition function zeroes of Ising models in quantum simulators}

\author{Abijith Krishnan}
\affiliation{Department of Physics, Harvard University, Cambridge, MA 02138, USA}
\author{Markus Schmitt}
\email{markus.schmitt@berkeley.edu}
\affiliation{Max-Planck-Institut f\"ur Physik komplexer Systeme, N\"othnitzer Stra{\ss}e 38,  01187 Dresden, Germany}
\affiliation{Department of Physics, University of California, Berkeley, CA 94720, USA}
\author{Roderich Moessner}
\affiliation{Max-Planck-Institut f\"ur Physik komplexer Systeme, N\"othnitzer Stra{\ss}e 38,  01187 Dresden, Germany}
\author{Markus Heyl}
\affiliation{Max-Planck-Institut f\"ur Physik komplexer Systeme, N\"othnitzer Stra{\ss}e 38,  01187 Dresden, Germany}
	
\begin{abstract}
Studying the zeroes of partition functions in the space of complex control parameters allows to understand formally how critical behavior of a many-body system can arise in the thermodynamic limit despite various no-go theorems for finite systems. In this work we propose protocols that can be realized in quantum simulators to measure the location of complex partition function zeroes of classical Ising models. The protocols are solely based on the implementation of simple two-qubit gates, local spin rotations, and projective measurements along two orthogonal quantization axes. Besides presenting numerical simulations of the measurement outcomes for an exemplary classical model, we discuss the effect of projection noise and the feasibility of the implementation on state of the art platforms for quantum simulation.
\end{abstract}
	
\maketitle

\section{Introduction}
The theory of phase transitions is central for our understanding of many-body systems.
However, formally explaining the singular behavior of thermodynamic potentials at phase transitions has been a challenging problem for a long time because the partition functions are generically analytic for systems of finite size~\cite{fisher1965}. 
One resolution of this problem is an artificial extension of system parameters to the complex plane.
In this framework, the nonanalytic properties of the thermodynamic potentials are  governed solely by the complex zeroes of the partition function, termed Fisher zeroes for complex temperatures~\cite{fisher1965} and Lee-Yang zeroes for complex fields~\cite{1952PhRv...87..404Y}.
While this extension has been primarily considered a mathematical tool, the interest in studying complex partition functions has recently been revived in different contexts, such as complex networks \cite{Krasnytska2015,Krasnytska2016}, real time evolution of quantum many-body systems \cite{2013PhRvL.110m5704H, Heyl2018}, protein folding \cite{Lee2013}, complex renormalization group flows \cite{Denbleyker2010,Denbleyker2011}, Bose-Einstein condensation \cite{Borrmann2000,vanDijk2015}, dynamical phase transitions in stochastic systems \cite{Flindt2013, Brandner2017}, and general studies of many-body systems in complex coupling space \cite{Jacobsen06,Wei2014,2013arXiv1304.6314B}. Remarkably, following the theoretical proposal in \cite{Wei2012} Lee-Yang zeroes for one-dimensional Ising chains have been measured for the first time in a nuclear magnetic resonance experiment~\cite{2015PhRvL.114a0601P}.

\begin{figure}
	\centering
	\includegraphics[width=\columnwidth]{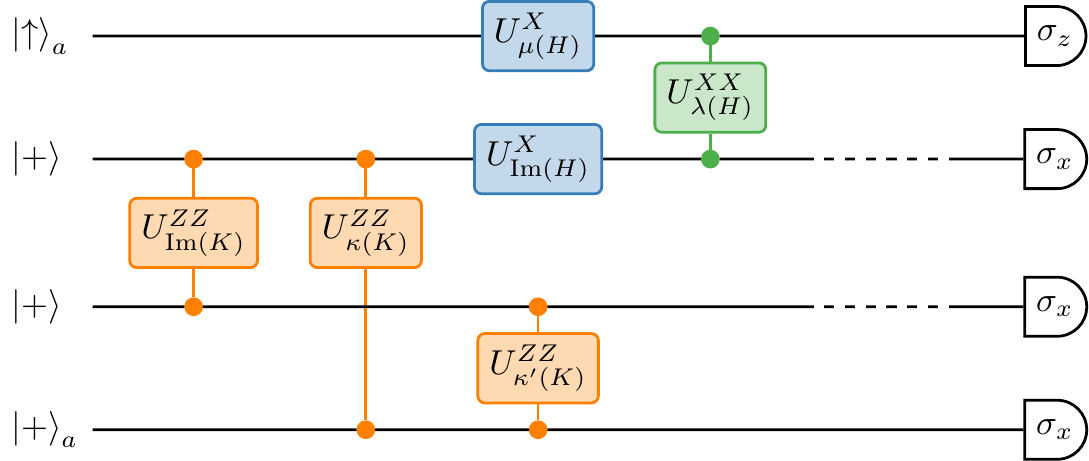}
	\caption{Basic building block of the quantum circuit that allows to measure complex partition function zeroes of a two-dimensional classical Ising model. The scheme involves unitary single-qubit and two-qubit gates. Ancilla bits $\ket{+}_a$ and $\ket{\uparrow}_a$ are required to explore the complex parameter plane and the dashed lines indicate further applications of the same sequence to the physical qubits, involving additional ancillas.}
	\label{fig:1}
\end{figure}

In this work, we show how the complex partition function zeroes for \textcolor{black}{a large class of} classical Ising models on various graphs and in different dimensions can be measured in quantum simulators such as trapped ions, superconducting qubits, or Rydberg atoms.
Our proposed experiment allows for a wide flexibility not only in the underlying geometry of the lattice but also in the different system parameters that can be extended to the complex plane.
On the experimental side, the protocol requires (i) the ability to initialize the qubits in a product state, (ii) the implementation of individual two-qubit Ising couplings, (iii) and the projective measurement of all the involved qubits in a fixed basis without the need of reconstructing the full quantum state.
In Fig.~\ref{fig:1}, we show an elementary building block of the proposed quantum circuit.
Full local control over the individual spin degrees of freedom is not necessary in order to realize interesting cases such as the two-dimensional Ising model, but it allows for a wider variety of realizable model systems.
We focus on Ising models, but the approach is equivalently applicable to Potts models.
In our proposed protocol, the number of qubits to map out the full complex plane for a parameter scales linearly with the simulated system size $N$.
In the end, we discuss the feasibility of the protocols in quantum simulation platforms such as trapped ions, superconducting qubits, and Rydberg atoms, as well as the influence of projection noise resulting from a finite number of measurements.

\section{Protocol}
We first present the experimental protocol before describing how it can be used to measure the partition function zeroes of Ising models.

We initialize the set of $M_P$ physical qubits in a product state
\be
	|\psi_{0,P} \rangle = |+\rangle = \bigotimes_{l=1}^{M_p} |+\rangle_l \, ,
\ee
which is the product of eigenstates \textcolor{black}{$\ket{+}_l$} of all the Pauli matrices $\sigma_l^x$, $l=1,\dots,M_p$, with eigenvalue $+1$.
For the most general case we will also require ancilla qubits, as we explain in detail later. We initialize these $M_A$ ancilla qubits, in a state $\ket{\psi_{0,A}}$, which can be either  polarized along the 
$z$- or $x$-axis. \textcolor{black}{The required number of ancilla qubits $M_A$ depends on the graph underlying the Ising Model, as we detail later in the text.}
Then our total initial state is 
\be
\ket{\psi_0} = \ket{\psi_{0,P}} \otimes \ket{\psi_{0,A}}.
\label{eq:ancilladef}
\ee
After the initialization we apply a sequence of two-qubit gates 
\be
	U^{\alpha\alpha}_\nu = e^{-i K \sigma_l^\alpha \sigma_m^\alpha} \, , \quad \nu=(l,m,K) \, ,
	\label{eq:gate1}
\ee
with $\sigma_l^\alpha$ a Pauli matrix ($\alpha=x,z$), $l,m$ denoting the two involved qubits, and $K$ the respective dimensionless coupling.
In addition, we also use local spin rotations:
\be
	U_\mu^\alpha = e^{-ih\sigma_l^\alpha} \, , \quad \mu = (l,h) \, .
    \label{eq:gate2}
\ee
After this sequence of unitary operations, the resulting state is of the form
\be
	|\psi \rangle = \bigotimes_{\mu} U_\mu^{\alpha_\mu} \, \bigotimes_{\nu} U_\nu^{\alpha_\nu\alpha_\nu} \, |\psi_0\rangle \, .
\ee
Finally, a projective measurement along the $\sigma^x$ axis is performed onto the initial condition:
\be
	\mathcal{L} = \big| \langle \psi_0 | \psi \rangle \big|^2 \,.
\ee
This measurement does not require full state tomography but can be estimated by the relative frequency at which the state $|+
\rangle$ appears in the projective measurement, as was done in recent experiments~\cite{2016Natur.534..516M,Jurcevic2017}.
As we show below, under a suitable choice of the gates the 
return probability \textcolor{black}{$\mathcal{L}$} is related to partition functions $Z$ of complex Ising models:
\be
	\mathcal{L} \propto |Z|^2 \, .
	\label{eq:Zprop}
\ee
Thus, zeroes of $Z$ are equivalently zeroes of \textcolor{black}{$\mathcal{L}$} and a measurement of \textcolor{black}{$\mathcal{L}$} provides the required information about Lee-Yang or Fisher zeroes of the implemented spin system.
When interested in not only the zeroes but also the full partition function $Z$, full state tomography is required to reconstruct the amplitude $\mathcal{G}(t) = \langle \psi_0 | \psi \rangle $ instead of the probability $\mathcal{L}(t)$.

\section{Partition function and return probability}
We consider the partition function of general classical Ising models including a magnetic field, which takes the form
\begin{align}
	Z&=\sum_{\vec s}\exp\Big(-\sum_{i,j}K_{ij}s_is_j-\sum_iH_is_i\Big)
	\nonumber\\&
	=\sum_{\vec s}\Bigg[\prod_{i,j}\exp\Big(-K_{ij}s_is_j\Big)\exp\Big(-H_is_i\Big)\Bigg]
	\label{eq:partitionSum}
\end{align}
Here, the sum runs over all possible configurations of $N$ spins, $\vec s\in\mathbb Z_2^N$, and $K_{ij}$ and $H_i$ denote dimensionless couplings or magnetic fields, respectively. Given that the operator $\exp\big(-\sum_{i,j}K_{ij}\sigma_i^z\sigma_j^z-\sum_iH_i\sigma_i^z\big)$ with $K_{ij} , H_i \in \mathbb{C}$ can be implemented, we directly obtain
\begin{align}
	\big|\braket{+|e^{-\sum_{i,j}K_{ij}\sigma_i^z\sigma_j^z-\sum_iH_i\sigma_i^z}|+}\big|^2
	=\frac{|Z|^2}{2^{2N}}\ ,
\end{align}
because $\ket{+}=\frac{1}{2^{N/2}}\sum_{\vec s}\ket{\vec s}$, where $\ket{\vec s}$ is the $\sigma^z$-basis.
Note that for this identity, we used the fact that the operator is diagonal in the $\sigma^z$-basis. Hence, the approach is restricted to classical partition functions. \textcolor{black}{Along the imaginary coupling axis, however, the partition function zeroes are related to Dynamical Quantum Phase Transitions for a quench from infinite to zero field of a quantum Ising model \cite{2013PhRvL.110m5704H,Heyl2015,Heyl2018}.}

As indicated in the second row of Eq.\ \eqref{eq:partitionSum}, the contributions of the interaction and field terms factorize such that these individual constituents can be considered separately. We additionally separate explicitly the unitary evolution given by the imaginary parts of the couplings, and the non-unitary part given by the real parts of the couplings. \textcolor{black}{Therefore, we expand our previous expression as}
\begin{align}
	 |\braket{+|\prod_{i,j}e^{-K_{ij}^R\sigma_i^z\sigma_j^z}e^{-i K_{ij}^I\sigma_i^z\sigma_j^z}e^{-H_I^R\sigma_i^z}e^{-iH_i^R\sigma_i^z}|+}|^2
\end{align}
with $K_{ij}=K_{ij}^R+iK_{ij}^I$ and $H_{j}=H_{j}^R+iH_{j}^I$\textcolor{black}{, and $K_{ij}^{R(I)},H_{ij}^{R(I)}\in\mathbb R$}. In this expression, the unitary parts directly correspond to the application of unitary gates $U_{(i,j,K_{ij}^I)}^{ZZ}$ and $U_{(i,H_{i}^I)}^{Z}$. As we demonstrate next, the non-unitary part can be implemented via the coupling to ancilla spins and additional projective measurements.

To emulate the action of the non-unitary operator $e^{-K_{ij}^R\sigma_i^z\sigma_j^z}$ on a state $\ket{\psi}$ with only unitary gates, we add an ancilla qubit to the system. This additional spin is polarized in the $x$-direction such that the state of the enlarged system is $\ket{\psi}\otimes\ket{+}_a$, \textcolor{black}{as alluded to in Eq.\ \eqref{eq:ancilladef}}. On this state, we apply two Ising gates \textcolor{black}{with coupling strength $\kappa_{ij}$ and $\kappa_{ij}'$ between qubits $i$, $j$ and the ancilla spin}, followed by a projection of the ancilla onto its initial state. \textcolor{black}{After using Eq.\ (3), this procedure yields
\begin{align}
	\bra{+}_aU_{(i,a,\kappa_{ij})}^{ZZ}U_{(j,a,\kappa_{ij}')}^{ZZ}\ket\psi\ket{+}_a
	=\cos(\kappa_{ij} \sigma_i^z + \kappa_{ij}' \sigma_j^z)  \ket{\psi}\ .
\end{align}}
The outcome can be matched with the effect of applying imaginary time evolution $e^{-K_{ij}^R\sigma_i^z\sigma_j^z}$ \textcolor{black}{by appropriately choosing $\kappa_{ij}$ and $\kappa_{ij}'$ so that the action of $\bra{+}_aU_{(i,a,\kappa_{ij})}^{ZZ}U_{(j,a,\kappa_{ij}')}^{ZZ}\ket\psi\ket{+}_a$ on basis configurations of spins $i$ and $j$ is equivalent to imaginary time evolution on those spins. Thus, after setting 
$\cos (2\kappa_{ij})= e^{-2|K_{ij}^R|}$ and $\kappa_{ij}'=\text{sgn}(K_{ij}^R)\kappa$, we obtain}
\begin{align}
	e^{-K_{ij}^R\sigma_i^z\sigma_j^z}\ket{\psi}=
	e^{|K_{ij}^R|}\bra{+}_aU_{(i,a,\kappa_{ij})}^{ZZ}U_{(j,a,\kappa_{ij}')}^{ZZ}\ket\psi\ket{+}_a\ .
\end{align}

Imaginary time evolution with a magnetic field is realized in the same fashion, again introducing an ancilla bit that is polarized along the $x$-axis. In this case, we require an Ising gate \textcolor{black}{with coupling $\lambda_j$} between the physical and the ancilla spin, and a local spin rotation \textcolor{black}{of strength $\mu_j$} on the ancilla, which results in
\begin{align}
	\bra{+}_aU_{(j,a,\lambda_j)}^{ZZ}U_{(a,\mu_j)}^Z\ket{\psi}\ket{+}_a
	=
	\cos(\lambda_j\sigma_j^z+\mu_j)\ket{\psi}\ .
\end{align}
\textcolor{black}{As with the previous paragraph, we adjust} the couplings $\lambda_j$ and $\mu_j$ to match the action \textcolor{black}{of $\bra{+}_aU_{(j,a,\lambda_j)}^{ZZ}U_{(a,\mu_j)}^Z\ket{\psi}\ket{+}_a$ with that of} the non-unitary operator by choosing \textcolor{black}{$\cos (2\lambda_j)=e^{-2\left|H^R_j\right|}$ and $\mu_j=-\text{sgn}(H^R_j)\lambda$}. Then, we obtain
\begin{align}
	e^{-H_j^R\sigma_j^z}\ket{\psi}
	=
	e^{|H_j^R|}\bra{+}_aU_{(j,a,\lambda_j)}^{ZZ}U_{(a,\mu_j)}^Z\ket{\psi}\ket{+}_a\ .
	\label{eq:longitudinal_field_gate}
\end{align}

\begin{figure*}[t!]
\includegraphics[width=.9\textwidth]{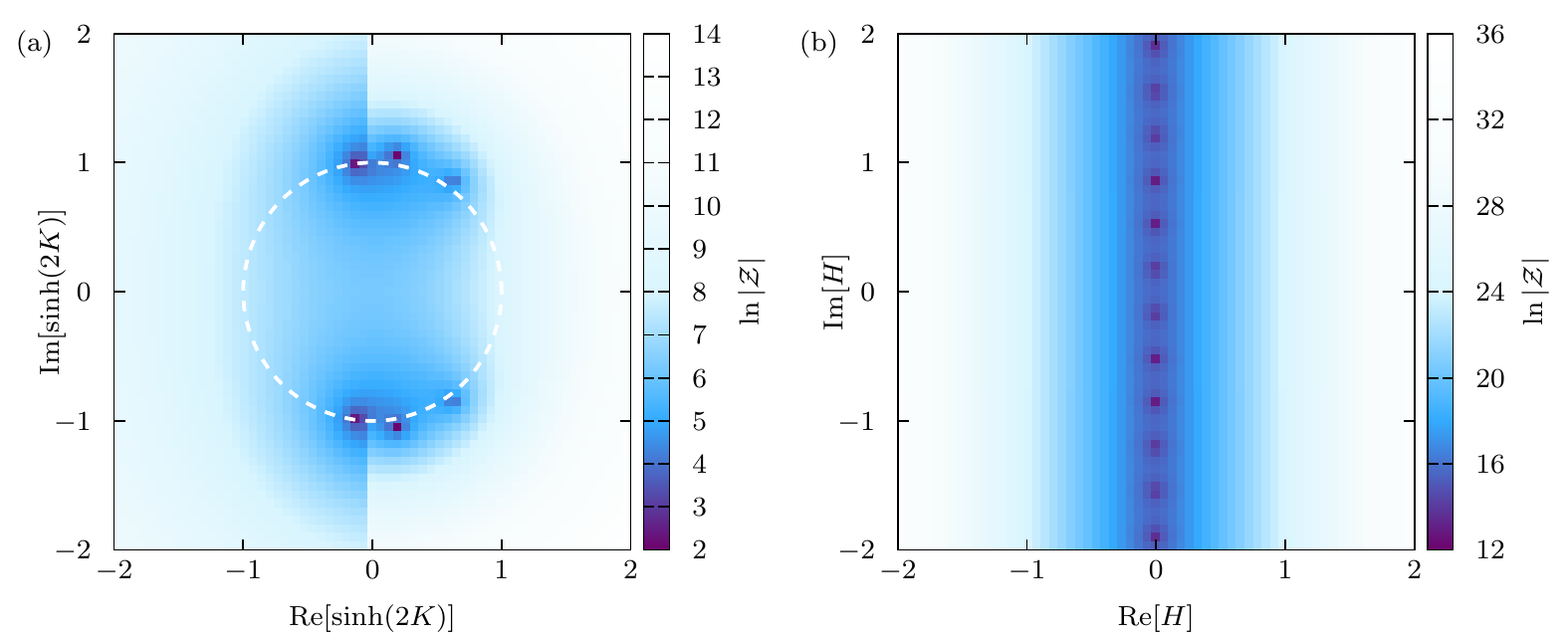}
\caption{
Complex partition function zeroes of a $3\times3$ Ising model on a cylinder as revealed by the return probability given in Eq.\ \eqref{eq:kicked_return_prob}.
\emph{(a)}
Fisher zeroes in the complex $K$-plane (note the rescaling of the axes). The white line indicates the unit circle, on which all Fisher zeroes reside in the thermodynamic limit.
\emph{(b)}
Lee-Yang zeroes in the complex magnetic field plane.
}
\label{fig2}
\end{figure*}
\textcolor{black}{After implementing our emulation of imaginary time evolution, our return probability $\mathcal{L}$ is 
\be 
\mathcal{L}=
	\big|\bra{\psi}_{0,A}\braket{+|U_a e^{-\sum_{i,j}iK_{ij}^I\sigma_i^z\sigma_j^z-\sum_jiH_j^I\sigma_j^z}|+}\ket{\psi}_{0,A}\big|^2\,
	\label{eq: returnprob}
\ee
where $\ket{\psi_{0,A}}$ is polarized in the $+x$ direction and
$U_A = \prod_{i,j}\left(U_{(i,a,\kappa_{ij})}^{ZZ}U_{(j,a,\kappa_{ij}')}^{ZZ}\right) \prod_j\left(U_{(j,a,\lambda_j)}^{ZZ}U_{(a,\mu_j)}^Z\right)$. Thus, we find the following} expression for $\mathcal{L}$:
\be
\mathcal{L} = \frac{e^{-2\sum_j|H_j^R| - 2\sum_j|K_{ij}^R|}}{2^{2N}}|Z|^2.
\ee
Thus, based on this scheme, one can obtain the norm of the partition function, $|Z|^2$, for arbitrary complex couplings $K_{ij}$ and $H_j$. This \textcolor{black}{procedure applies to}, in principle, Ising models on arbitrary graphs (although the concrete feasible realizations can depend on the experimental platform), meaning in particular that the procedure is independent of the dimensionality of the classical Ising model. 

In this approach, we introduced one qubit per classical spin. To map out, for example, the Fisher zeroes in the complex parameter plane of an Ising model without magnetic field on a square lattice with cylindric boundary conditions and lattice dimensions $N\times L$ (with open boundary of length $L$), $3NL-N$ qubits are required and $6NL-3N$ unitary gates need to be applied. The number of qubits arises from $NL$ ``physical" qubits and $2NL-N$ ancillas, which are necessary to realize the real parts of the Ising coupling. \textcolor{black}{We require $6NL - 3N$} unitary gates because for each Ising coupling, one gate is required for the real part and two gates are required for the imaginary part.

\section{Two-dimensional Ising models}
We now discuss an experimental protocol for two-dimensional Ising models, which requires fewer qubits \textcolor{black}{than the above procedure} and realizes open boundary conditions in one direction and open or periodic boundary conditions along the other direction.
For simplicity of notation, we assume homogeneous nearest neighbor Ising couplings $K_x$ and $K_y$ in $x$ and $y$ directions, respectively, and we exclude the longitudinal magnetic field, which can be implemented just as in the previous section; these constraints can be relaxed, however, as \textcolor{black}{our procedure can treat more general models analogously}.
\textcolor{black}{In our protocol, similar to the conventional quantum-classical mapping, we relate the return probability of a periodically kicked transverse field quantum Ising Model,
\be
P_{\text{kicked}}=|\braket{+|e^{-K\sum\sigma_i^z\sigma_j^z}\big(e^{-H\sum\sigma_i^x}e^{-K\sum\sigma_i^z\sigma_j^z}\big)^{L-1}|+}|^2\ ,
\ee 
to the partition function of a two-dimensional classical Ising Model after insertion of a complete set of states after each period.}
We then arrive at
\begin{align}
    P_{\text{kicked}}
	=\left|\frac{\sinh(2H)^{N(L-1)}}{2^{N(L+1)}}\right| \big|Z(K,-\frac{\ln(\tanh(H))}{2},N,L)\big|^2
\end{align}
where $Z(K_x,K_y,N,L)$ is the partition function of the classical Ising model with lattice size $N\times L$.
In the expression above, the quantum couplings $K$ and $H$ can again be complex numbers and the non-unitary part can be implemented as discussed in the previous paragraph. We implement the imaginary time evolution of the transverse field of the quantum model analogously to the imaginary time propagation $e^{-H_j^R\sigma_j^z}$ given in Eq.\ \eqref{eq:longitudinal_field_gate}, i.e., the coupling to a transverse field can be implemented as
\begin{align}
	e^{-H^R\sigma_j^x}\ket{\psi}
	=
	e^{|H^R|}\bra{\uparrow}_aU_{(j,a,\lambda)}^{XX}U_{(a,\mu)}^X\ket{\psi}\ket{\uparrow}_a\ .
\end{align}
with $\lambda$ and $\mu$ as given above and $U_{(j,a,\lambda)}^{XX}$ and $U_{(a,\mu)}^X$ are defined analogously to the gates introduced in Eqs.\ \eqref{eq:gate1} and \eqref{eq:gate2} with $\sigma^x$ Pauli operators. Here, $\ket{\uparrow}$ denotes the eigenstate of $\sigma^z$ with eigenvalue $+1$, $\sigma^z\ket{\uparrow}=\ket{\uparrow}$. Then, \textcolor{black}{after computing $\mathcal{L}$ analogously to Eq.\ \eqref{eq: returnprob} and, as discussed above, adding transverse field gates and $z$-polarized ancilla spins,} our expression for the measured probability $\mathcal{L}$ is 
\begin{align}
	\mathcal L = &\frac{|\braket{+|e^{-K\sum\sigma_i^z\sigma_j^z}\big(e^{-H\sum\sigma_i^x}e^{-K\sum\sigma_i^z\sigma_j^z}\big)^{L-1}|+}|^2}{e^{2N(L-1)|H^R| + 2NL|K^R|}}.
	\label{eq:kicked_return_prob}
\end{align}

\textcolor{black}{In Fig.\ \ref{fig:1}, we display} an example of the basic building block of the corresponding quantum circuit that implements the required kicking. The quantum circuit involves the Ising coupling between two qubits and the evolution of single qubits in the transverse field. For the non-unitary part of the evolution, two ancilla bits are required.

Overall, to address a classical Ising model on a square lattice with cylindric boundary conditions of size $N\times L$, $2NL$ qubits are required as opposed to $3NL-N$ qubits that would be required using the scheme of the previous paragraph. Although the number of qubits is still proportional to the number of classical spins, the smaller prefactor is relevant considering the number of qubits available in present day quantum simulators. \textcolor{black}{Note that the number of ancilla qubits could be further reduced on platforms that allow for the measurement and re-initialization of individual qubits during the execution of a circuit.} The total number of required gates is $6NL-3N$, just as in the scheme presented before.

\section{Complex partition function zeroes}
To identify zeroes of the partition function in the complex parameter plane experimentally, one  maps out the parameter space with measurements of the return probability $\mathcal L$ as described above. We now present hypothetical results obtained from numerical simulations of the experimental protocols based on an exact numerical simulation. For these examples we chose a two-dimensional Ising model on a square lattice of $3\times3$ spins with open boundary conditions in one direction and periodic boundaries in the other direction for convenience. We consider identical couplings of all pairs of nearest neighbors. For this setting, the realization appears within reach using the kicking protocol \eqref{eq:kicked_return_prob} with state of the art quantum simulators.
First, we consider Fisher zeroes, i.e., roots of the partition function in the complex $K$-plane. The numerical result for 
the partition function $Z$ as given in Eq.\ \eqref{eq:kicked_return_prob} is shown in Fig.\ \ref{fig2}(a). In the complex coupling plane, the result shows distinct points of a vanishing partition function corresponding to Fisher zeroes of $Z$. Upon increasing the system size the number of Fisher zeroes also increases. In the thermodynamic limit, they coalesce to form manifolds in the complex plane~\cite{fisher1965}.
In the case shown here, this manifold is the unit circle \cite{fisher1965,vanSaarloos1983,Lu2001}, indicated as the white dashed line in Fig. \ref{fig2}(a).
In Fig.\ \ref{fig2}(b), we show the 
partition function $Z$ in the complex magnetic field plane to identify Lee-Yang zeroes. Again, the return probability shows distinct points at which $Z$ becomes zero. The Lee-Yang zeroes lie on the imaginary axis; thus, there is no phase transition as a function of the magnetic field for $\text{Re}(H)>0$.

\begin{figure}[t!]
\includegraphics[]{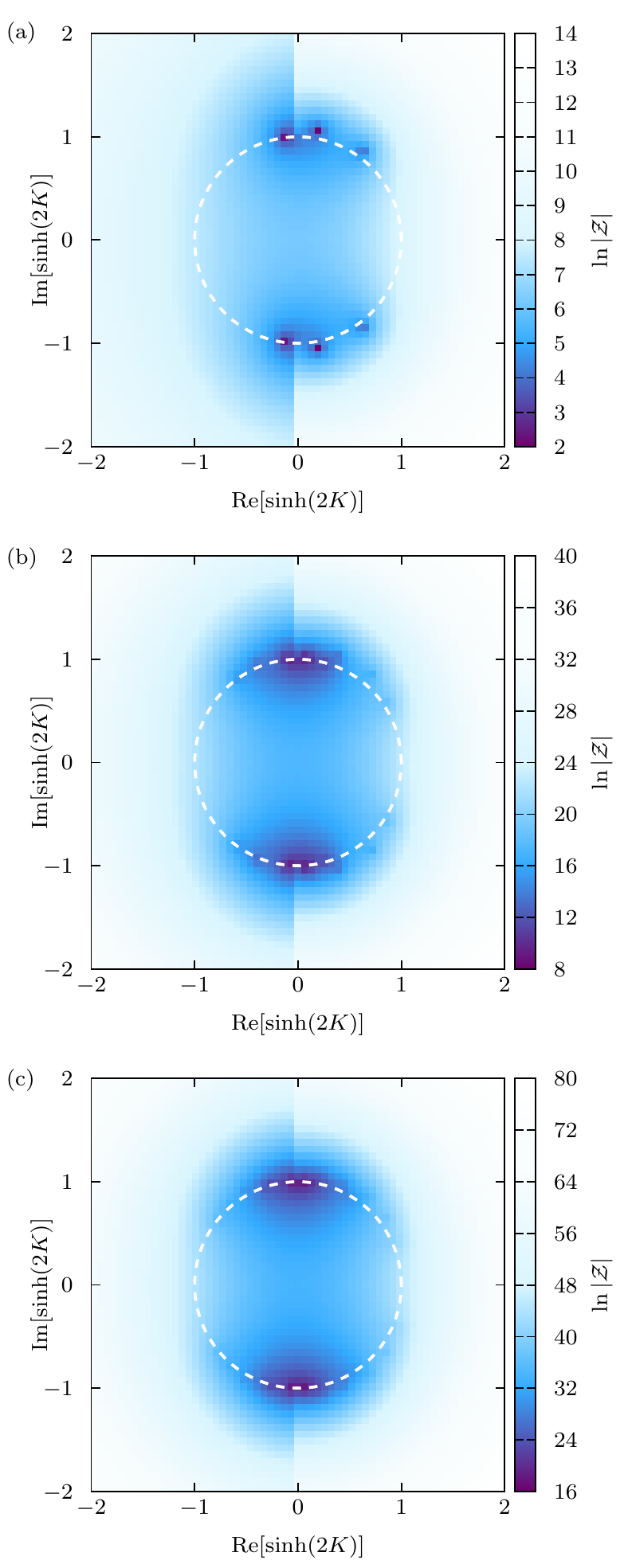}
\caption{
Complex partition function zeroes of Ising models on cylinders of different sizes: (a) $3\times3$, (b) $5\times5$, (c) $7\times7$. With increasing system size the distribution of Fisher zeroes becomes denser. In the thermodynamic limit they will coalesce on the unit circle indicated by the white dashed line.
}
\label{fig:appendix}
\end{figure}

As one can see, the complex zeroes for the small system considered in Fig.\ \ref{fig2}(a) already follow closely the result in the thermodynamic limit.
In Fig.\ \ref{fig:appendix}, we show the distribution of Fisher zeroes in the complex temperature plane of an Ising model with cylindric boundary conditions for system sizes $3\times3$, $5\times5$, and $7\times7$.
The number of zeroes increases with increasing system size, and they are distributed more and more densely along the unit circle indicated by the white dashed lines. In the thermodynamic limit the Fisher zeroes will coalesce to this unit circle that cuts the real parameter axis, thereby indicating the critical temperature of the phase transition between ferromagnet and paramagnet. The density of partition function zeroes near the critical temperature determines critical exponents of the transition~\cite{fisher1965}. Therefore, the study of partition function zeroes for finite systems can already give information about the thermal phase transition.

\begin{figure}[t!]
\includegraphics[width=\columnwidth]{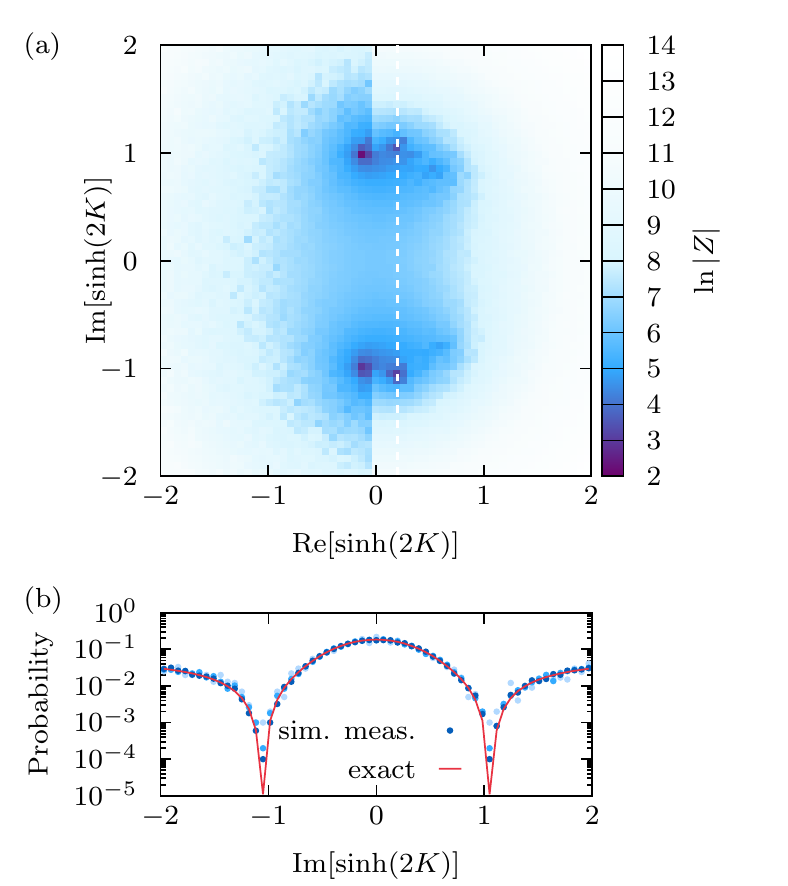}
\caption{(a) Simulated measurement outcome for the partition function zeroes in the complex coupling plane of a classical $3\times3$ Ising model for a finite number of 5000 measurements displaying the projection noise. (b) Cut along the white dashed line of (a) comparing the measured probability $\mathcal{L}$ with the exact result. From light to dark blue the data points correspond to simulated measurements with 1000, 5000, and 10000 samples.
}
\label{fig3}
\end{figure}
\section{Projection noise}
In the experimental realization of the proposed protocols the return probability $\mathcal L$ will be obtained from a finite number of measurements as the relative frequency of the outcome that all spins are polarized along the positive $x$-direction.
This finite number of measurements introduces projection noise onto the result for $\mathcal L$. In order to assess whether partition function zeroes can still be identified, we simulate numerically the effect of projection noise by drawing a finite number of samples from the resulting wave function.

In Fig. \ref{fig3} a representative result is shown for the measured partition function, obtained with $N=5000$ samples at each point in the complex coupling plane. Despite the fluctuations, the locations of Fisher zeroes can be identified as distinct minima in the observed return probability.

Having discussed the possibilities to detect experimentally the location of partition function zeroes in the complex parameter plane, we next extend this scheme to the measurement of correlation functions.

\section{Correlation functions}
In thermal ensembles correlation functions of the form
\begin{align}
	\braket{s_is_j}=\frac{1}{Z}\sum_{\vec s}s_is_je^{-\sum_{lm}K_{lm}s_ls_m-\sum_lH_ls_l}
\end{align}
probe physical properties of the system, whose extensions to complex parameter planes have been also explored~\cite{2013arXiv1304.6314B}. As we explain now, in addition to the partition function, accessing correlation functions at complex couplings is also possible within the proposed framework of quantum circuits. In particular, the norm of the correlation function is accessible by slightly extending the scheme discussed before for the measurement of the partition function.
To implement the measurement of correlation functions in the quantum circuit we exploit the fact that $\sigma_i\sigma_j=-ie^{i\pi \sigma_i\sigma_{j-1}/2}$. Therefore, the norm of the correlation function can be obtained as the ratio the return probabilities in two quantum circuits that differ only by one additional Ising gate,
\begin{align}
	|\braket{s_is_j}|^2=
	\frac{\big|\braket{+|e^{i\pi/2(\sigma_i\sigma_j-1)-\sum_{i,j}K_{ij}\sigma_i^z\sigma_j^z-\sum_iH_i\sigma_i^z}|+}\big|^2}{\big|\braket{+|e^{-\sum_{i,j}K_{ij}\sigma_i^z\sigma_j^z-\sum_iH_i\sigma_i^z}|+}\big|^2}\ .
\end{align}

Now, we present a procedure for computing the correlation function for spins which includes information about phases, in the two-dimensional Ising Model. This method can also be generalized to arbitrary Ising models. There are two mechanisms for explicitly computing complex correlation functions between two spins for the two-dimensional Ising Model -- one for both spins in the same ``row'' of the lattice and one for the case where the spins are located in different rows. Both involve inserting additional couplings or magnetic fields at the spins whose correlation we wish to measure. 

Consider our experimental procedure including kicking for the two-dimensional Ising Model with couplings $K$ and $H$. To measure correlations between spins $(i,j)$ and $(k,l)$ one can introduce an additional coupling $\delta K_{ij,kl}$ between those spins. For this section, we let $\mathcal F(\delta K_{ij,kl})$ be the return amplitude measured for the two-dimensional Ising Model with said additional coupling.

First, suppose the spins are in the same row, i.e., both spins occur in the same kick of the kicked quantum Ising Model. We first add a small real coupling $\delta K^R$ between the two spins whose correlation we are measuring: $S_{i,m}$ and $S_{k,m}$. Then, the magnitude of the return amplitude is given by 
\be
|\mathcal F(\delta K_{im,km}^R)|^2 = \frac{|\mathcal F(0)|^2(1+ 2 \delta K_{im,km}^R\text{Re} \langle{S_{i,m}S_{k,m}}\rangle)}{e^{2(|K+\delta K^R_{im,km}|-|K|)}}.
\ee
If we instead add a small imaginary coupling $\delta K_{im,km}^I$ between the two spins $S_{i,m}$ and $S_{k,m}$, we get 
\be
|\mathcal F(\delta K_{im,km}^I)|^2 =|\mathcal F(0)|^2(1- 2 \delta K_{im,km}^I\text{Im} \langle{S_{i,m}S_{k,m}}\rangle).
\ee
Therefore, making measurements with an additional real or imaginary coupling between the spins of interest, respectively, will provide real and imaginary parts of $\langle S_{i,m} S_{k,m}\rangle$ as
\begin{align}
    \text{Re} \langle{S_{i,m}S_{k,m}}\rangle
    &= -  \frac{1}{2\delta K_{im,km}^R} \nonumber \\
    &\quad+\frac{|\mathcal F(\delta K_{im,km}^R)|^2 e^{2(|K+\delta K^R_{im,km}|-|K|)}}{2\delta K_{im,km}^R|\mathcal F(0)|^2 }
    \nonumber\\
    \text{Im} \langle{S_{i,m}S_{k,m}}\rangle
    &=
    -\frac{1}{2\delta K_{im,km}^I}\Bigg(\frac{|\mathcal F(\delta K_{im,km}^I)|^2}{|\mathcal F(0)|^2}-1\Bigg).
\end{align}

Now, suppose the spins are in different rows. Because the spins occur during different kicks of the kicked quantum Ising Model, coupling the spins is not feasible. We thus add $z$-direction magnetic fields at each spin: $S_{i,m}$ and $S_{k,n}$. We let $\mathcal F(\delta B_{ij,kl})$ be the return amplitude measured for the two-dimensional Ising Model with an additional $z$-direction magnetic field $\delta B_{ij,kl}$ added at both spins $ij$ and $kl$. First, we add a real magnetic field $\delta B^R_{im,kn}$ to both spins.  Then,
\begin{align}
|Z(\delta B^R_{im,kn})|^2 =|Z(0)|^2(1+ 2(\delta B^R_{im,kn})^2(1 +\nonumber\\ \text{Re} \langle S_{i.m}S_{k,n}\rangle))
\end{align}
and
\begin{align}
&\mathcal{F}(\delta B^R_{im,kn})=
\nonumber\\
&e^{-4|B^R_{im,kn}|} \mathcal{F}(0)(1+ 2(\delta B^R_{im,kn})^2(1 + \text{Re} \langle S_{i.m}S_{k,n}\rangle))\ .
\end{align}
Thus, measuring $\mathcal{F}(\delta B^R_{im,kn})$ will indicate the real part of $\langle S_{i,m}S_{k,n} \rangle$. Now, we add a magnetic field $\delta B_{im,kn}^R e^{i\pi/4}$ at each spin. Then, 
\begin{align}
|Z(\delta B^R_{im,kn}e^{i\pi/4})|^2 =|Z(0)|^2(1- 2(\delta B^R_{im,kn})^2\cdot\nonumber\\ \text{Im} \langle S_{i.m}S_{k,n}\rangle).
\end{align}
and
\begin{align}
&\mathcal{F}(\delta B^R_{im,kn} e^{i\pi/4})=
\nonumber\\
&e^{-2\sqrt{2}\delta B^R_{im,kn}}\mathcal{F}(0)(1- 2(\delta B^R_{im,kn})^2\text{Im} \langle S_{i.m}S_{k,n}\rangle)\ ,
\end{align}
which yields the imaginary part of the desired classical expectation value. Thus, both a simpler procedure for measuring the norm of a correlation function and a more involved procedure for measuring the phase of a correlation function are possible with the quantum circuit presented in this paper. 

\section{Discussion}
In this work, we have proposed schemes to measure complex partition function zeroes of arbitrary Ising models in quantum simulators. The required number of qubits scales linearly in the system size of the target system, where the details of the scaling depend on the chosen boundary conditions and on the specific simulation scheme.

For Ising models of small size, our scheme is feasible on current quantum simulation platforms including trapped ions, superconducting qubits, or Rydberg atoms, as we will argue in the following. In all of these devices, the simulation of more than $10$ qubits has been reported~\cite{2017Natur.551..579B,2017arXiv171205771O,2018PhRvX...8b1070L,Kokail2019}, initial product states have been realized~\cite{2017Natur.551..579B,2017arXiv171205771O,2018PhRvX...8b1070L,Kokail2019,Jurcevic2017,Zhang2017, 2017Natur.551..579B,Levine2018,Leseleuc2018}, the individual gates can be implemented, and single-qubit resolved measurements in a fixed measurement basis can be performed \cite{2017Natur.551..579B,2017arXiv171205771O,2018PhRvX...8b1070L,Kokail2019,Jurcevic2017,Zhang2017, 2017Natur.551..579B,Levine2018,Leseleuc2018}. For a concrete choice of Ising model, coupling the ancilla qubits to the physical ones properly may be challenging. Using trapped ion or superconducting qubit quantum computers any desired Ising coupling between two individual spins can be achieved, in principle. While controlled experiments with 20 or more qubits remain challenging, substantial progress has been reported recently~\cite{2017arXiv171205771O,Zhang2017,2017Natur.551..579B,Kokail2019,Keesling2019,Kandala2019}.

\textcolor{black}{
Let us now outline more concretely one potential realization in systems of Rydberg atoms. Targeting the measurement of the partition function for a one-dimensional Ising model, for instance, one can utilize} the recent advances to controllably place Rydberg atoms in three-dimensional space by means of optical tweezers \cite{Barredo2016,2017Natur.551..579B,Barredo2018}. Starting with a linear chain to realize the targeted Ising model, the ancilla spins can be placed midway between two physical spins but outside of the one-dimensional line forming the Ising chain. \textcolor{black}{What remains is to engineer the spin-spin interactions between the spins of the targeted model and their coupling to the ancillas. Here, one can make use of two degrees of freedom to tune the involved interaction strengths to their desired values. On the one hand, the interaction strength exhibits a marked distance dependence which can take either dipolar or van-der Waals form. On the other hand,  interactions also show a directional dependence due to their dipolar nature~\cite{2018arXiv181013286D}. The combination of those two ways to tune the involved interaction strengths allows to explore a wide range of complex parameter space. For a chain of length $N=10$, this would require a total number of $19$ spins which is even well below reported experiments with systems realizing more than $50$ spins~\cite{2017Natur.551..579B,Keesling2019}.}

While the scheme relies on the simulation of only a number of qubits linear in the system size of the targeted system, the measurement of the zeroes requires nevertheless resources scaling exponentially. This increase in scale, however, is not a consequence of our proposed protocols but rather due to the fact that partition functions depend exponentially on system size.

\textcolor{black}{Previously, the calculation of partition functions for classical spin models has been related to quantum computation in ways that, however, differ from our approach. These relations allow to use techniques from quantum information theory to compute the partition sum on classical computers \cite{vanDenNest2007}, to draw conclusions about the classical simulability of quantum algorithms from the knowledge of classical partition sums \cite{vanDenNest2009}, or to construct efficient quantum algorithms to evaluate the partition sum of $\pm J$ spin glasses \cite{Lidar2004}. By contrast, our scheme can, in principle, translate the complex partition function of any classical Ising model on any graph to a circuit for execution on a quantum simulator.} Further, the protocols can be straightforwardly extended to classical Potts models.

\begin{acknowledgments}
Support by the Deutsche Forschungsgemeinschaft via the Gottfried Wilhelm Leibniz Prize program is gratefully acknowledged. MS was supported through the Leopoldina Fellowship Programme of the German National Academy of Sciences Leopoldina (LPDS 2018-07) with additional support from the Simons Foundation. AK was supported through the Harvard University Herchel Smith Fellowship, administered by the Office of Undergraduate Research and Fellowships, as well as through the Visitors Program of the Max Planck Institute for the Physics of Complex Systems.
\end{acknowledgments}

\bibliography{literature.bib}

\end{document}